\documentclass[reqno,11pt]{amsart}

\IfFileExists{mymtpro2.sty}{%
  \usepackage[subscriptcorrection]{mymtpro2}}%
{%
  \let\wtilde\widetilde%
}

\usepackage{a4,amssymb}

\numberwithin{equation}{section}

\newtheorem{theorem}{Theorem}[section]

\theoremstyle{remark}
\newtheorem{remark}[theorem]{Remark}
\newtheorem{myremarks}[theorem]{Remarks}

\newenvironment{acknowledgement}{\par\medskip\noindent\emph{Acknowledgement.}}

\newenvironment{remarks}{\begin{myremarks}\begin{nummer}}%
    {\end{nummer}\end{myremarks}}

\newcounter{numcount}
\newcommand{\labelnummer}{\mbox{\normalfont (\roman{numcount})}}%

\makeatletter

\newenvironment{nummer}%
  {\let\curlabelspeicher\@currentlabel%
    \begin{list}{\labelnummer}%
      {\usecounter{numcount}\leftmargin0pt%
        \topsep0.5ex\partopsep2ex\parsep0pt\itemsep0ex\@plus1\p@%
        \labelwidth2.5em\itemindent3.5em\labelsep1em%
      }%
    \let\saveitem\item%
    \def\item{\saveitem%
      \def\@currentlabel{{\upshape\curlabelspeicher}$\,$\labelnummer}}%
    \let\savelabel\label%
    \def\label##1{\savelabel{##1}%
      \@bsphack%
        \ifmmode\else%
          \protected@write\@auxout{}%
          {\string\newlabel{##1item}{{\labelnummer}{\thepage}}}%
        \fi%
      \@esphack%
    }%
  }{\end{list}}%

\renewcommand\d{\mathrm d}
\newcommand\e{\mathrm{e}}

\def\le{\leqslant} \let\leq\le
\def\ge{\geqslant} \let\geq\ge

\def\Chi{\raisebox{.4ex}{$\chi$}}

\DeclareMathOperator{\tr}{tr\kern1pt}

\makeatletter

\newif\ifper\pertrue
\def\per{.}

\def\au#1#2{#2, #1}
\def\lau#1#2{#2, #1:}

\def\ti#1{#1\ifper.\fi\pertrue}

\def\bti{\@ifnextchar[\bbti\bbbti}
\def\bbti[#1]#2{#2, #1.}
\def\bbbti#1{#1.}

\def\z{\@ifnextchar[\zz\zzz}
\def\zz[#1]#2#3#4#5{\perfalse\emph{#2} \textbf{#3}, #4 (#5) [#1]}
\def\zzz#1#2#3#4{\emph{#1} \textbf{#2}, #3 (#4)\ifper\per\fi\pertrue}

\def\pub{\@ifstar\pubstar\pubnostar}
\def\pubnostar{\@ifnextchar[\@@pubnostar\@pubnostar}
\def\@@pubnostar[#1]#2#3#4{#2, #3, #4, #1\ifper\per\fi\pertrue}
\def\@pubnostar#1#2#3{#1, #2, #3\ifper\per\fi\pertrue}
\def\pubstar[#1]#2#3#4{\perfalse #2, #3, #4 [#1]\pertrue}

\makeatother

\newcommand{\beq}{\begin{equation}}
\newcommand{\eeq}{\end{equation}}
\newcommand{\ba}{\begin{array}}
\newcommand{\ea}{\end{array}}
\newcommand{\bea}{\begin{eqnarray}}
\newcommand{\eea}{\end{eqnarray}}

\newcommand{\R}{\mathbb{R}}
\newcommand{\Rd}{\mathbb{R}^d}

\newcommand{\Z}{\mathbb{Z}}

\newcommand{\N}{\mathbb{N}}
\newcommand{\NN}{\mathbb{N}}

\newcommand{\Q}{\mathbb{Q}}

\newcommand{\cU}{\mathcal{U}}
\newcommand{\cH}{\mathcal{H}}
\newcommand{\cK}{\mathcal{K}}
\DeclareMathOperator{\supp}{\mathrm{supp}}

\def\P{I\kern-.30em{P}}
\def\E{I\kern-.30em{E}}
\renewcommand{\E}{\mathbb{E}\mkern2mu}
\renewcommand{\P}{\mathbb{P}}

\begin{document}

\title[The spectral shift function] {The spectral shift function for
  compactly supported perturbations of Schr\"odinger operators on large bounded domains}

\author[P.\ D.\ Hislop]{Peter D.\ Hislop}
\address{Department of Mathematics,
    University of Kentucky,
    Lexington, Kentucky  40506-0027, USA}
\email{hislop@ms.uky.edu}

\author[P.\ M\"uller]{Peter M\"uller}
\address{Mathematisches Institut,
  Ludwig-Maximilians-Universit\"at,
  Theresienstra\ss{e} 39,
  80333 M\"unchen, Germany}
\email{mueller@lmu.de}

\thanks{Appeared in: \z{Proc.\ Amer.\ Math.\ Soc.}{138}{2141--2150}{2010}}

\begin{abstract}
  We study the asymptotic behavior as $L \rightarrow \infty$ of the
  finite-volume spectral shift function for a positive,
  compactly-supported perturbation of a Schr\"odinger operator in
  $d$-dimensional Euclidean space, restricted to a cube of side length
  $L$ with Dirichlet boundary conditions. The size of the support of
  the perturbation is fixed and independent of $L$.  We prove that the
  Ces\`aro mean of finite-volume spectral shift functions remains
  pointwise bounded along certain sequences $L_n \rightarrow \infty$
  for Lebesgue-almost every energy.  In deriving this result, we give
  a short proof of the vague convergence of the finite-volume spectral
  shift functions to the infinite-volume spectral shift function as $L
  \rightarrow \infty$. Our findings complement earlier results of W.\
  Kirsch [Proc.\ Amer.\ Math.\ Soc.\ {\bf 101}, 509--512 (1987), Int.\
  Eqns.\ Op.\ Th.\ {\bf 12}, 383--391 (1989)] who gave examples of
  positive, compactly-supported perturbations of finite-volume
  Dirichlet Laplacians for which the pointwise limit of the spectral
  shift function does not exist for any given positive energy. Our
  methods also provide a new proof of the Birman--Solomyak formula for
  the spectral shift function that may be used to express the measure
  given by the infinite-volume spectral shift function directly in
  terms of the potential.
\end{abstract}

\maketitle
\thispagestyle{empty}


\section{Statement of the Problem and Result}

The spectral shift function (SSF) plays an important role in scattering theory
for Schr\"odinger operators \cite{Yaf92}. 
For the particular case of a quantum mechanical system in a finite volume, the
SSF, as a function of energy $E$, counts the change in the number of
eigenvalues below $E$ due to adding a perturbing potential $V$.

We are interested in the following question: given cubes $\Lambda_L \subset
\R^d$ in $d$-dimensional Euclidean space, which are centered at the origin and
have side lengths $L > 0$, what is the limiting behavior as $L\to\infty$ of
the SSF corresponding to the Laplacian plus a background potential
$H_{0}^{(L)} := - (\Delta_L / 2) + V_{0}$ and its perturbation $H_1^{(L)} :=
H_0^{(L)} + V$? Both operators are defined on the Hilbert space
$\mathrm{L}^{2}(\Lambda_{L})$ with Dirichlet boundary conditions.  The
potentials $V_{0}$ and $V$ act as multiplication operators corresponding to
real-valued functions $V_{0}$ and $V$ such that
\begin{equation}
  \label{ass}
  \tag{$\star$}
  \begin{split}
    &\max\{0, V_{0}\} \in \cK_{\mathrm{loc}}(\R^{d}), \quad \max\{0, -V_{0}\}
    \in \cK(\R^{d}), \\
    &V \in\cK_{\mathrm{loc}}(\R^{d}) , \quad V \ge 0, \quad \supp (V)
    \text{~ is compact}.
  \end{split}
\end{equation}
It is understood in \eqref{ass} that the compact support of $V$ is independent
of $L$, and we have written $\mathcal{K}(\R^{d})$ and
$\mathcal{K}_{\mathrm{loc}} (\R^d)$ to denote the Kato class and the local
Kato class, respectively \cite{AiSi82, Sim82}. We also introduce the
corresponding infinite-volume self-adjoint Schr\"odinger operators $H_{0} :=
-(\Delta/2) + V_{0}$ and $H_{1}:= H_{0} + V$ on $\mathrm{L}^{2}(\R^{d})$.

The self-adjoint operators $H_{0}^{(L)}$ and $H_{1}^{(L)}$ have compact
resolvents and, therefore, discrete spectrum. For a given energy $E \in \R$,
let $N_0^{(L)}(E)$, resp.\ $N_1^{(L)}(E)$, denote the number of eigenvalues,
including multiplicity, for $H_0^{(L)}$, resp.\ $H_1^{(L)}$, less than or
equal to $E$.  These are both monotone increasing functions of the energy $E$.
We define the \emph{relative eigenvalue counting function} by
\beq\label{ssf2}
  E \mapsto \xi_L (E) \equiv \xi (E; H_1^{(L)}, H_0^{(L)})
  := N_0^{(L)}(E) -N_1^{(L)}(E) \geq 0
\eeq
for all $E \in \R$.  It is known that this function is equal to the (more
generally defined) \emph{spectral shift function} for the pair $(H_1^{(L)},
H_0^{(L)})$, see e.g.\ \cite{Yaf92, BiYa93} or \eqref{ssf-def-0} in the
Appendix.  A basic question is the pointwise boundedness with respect to the
energy of the SSF $\xi_L$ as $L \rightarrow \infty$.

The main result of this note is Theorem~\ref{kom-cor} which states that the
Ces\`aro mean of subsequences of $\xi_{L}$ is bounded from above by the
infinite-volume SSF $\xi$ Lebesgue-almost everywhere. Here, the
infinite-volume SSF $\xi$ for the pair $(H_{1}, H_{0})$ is defined in terms of
the invariance principle and Kre\u{\i}n's trace identity, see
Remark~\ref{rem-ssf} in the Appendix.  Theorem~\ref{kom-cor} relies on an
abtract result of Koml\'os \cite{Kom67} and vague convergence of the measures
$\xi_{L}(E) \,\d E$ to the measure $\xi(E) \,\d E$, where $\d E$ denotes the
Lebesgue measure on $\R$.  As a by-product of our analysis we obtain a 
short proof of the Birman--Solomyak formula for the SSF of Schr\"odinger
operators that seems to be new.

The main motivation of this note are the two papers of W.\ Kirsch \cite{Kir87,
  Kir89}, who considered the (un-)boundedness of the SSF $\xi_L$ as $L
\rightarrow \infty$. Weyl's law indicates that the leading behavior of each
eigenvalue counting function in (\ref{ssf2}) is the same and proportional to
the volume $L^d$. Since the support of $V$ is compact and independent of $L$,
one might think that all $L$-dependence in (\ref{ssf2}) cancels out and that
the SSF remains bounded as $L \rightarrow \infty$. For the corresponding
discrete problem in $\ell^{2}(\Z^d)$, this is indeed true as can be seen from
a finite-rank perturbation argument. For the continuum problem, however, which
we consider here, Kirsch showed that this intuition is wrong in dimensions $d
\geq 2$ if $ V_{0} \equiv 0$. (In $d=1$ the finite-rank perturbation argument
is also applicable in the continuum.)

\begin{theorem}[\cite{Kir87}] \label{kirsch1} Let $d \in \NN\setminus \{1\}$
  and assume in addition
  to \eqref{ass} that $V_{0}=0$ and $V \in
  \mathrm{L}^{\infty}(\Rd)$. Then, for any $E > 0$, we have
  \beq\label{unbdd1}
    \sup_{L > 0} \xi_L (E) = \infty .
  \eeq
  Furthermore, there is a countable, dense set of
  energies $\mathcal{E} \subset [0, \infty[$ so that for any $E \in
  \mathcal{E}$, we have
  \beq\label{unbdd2}
    \sup_{L \in \N} \xi_L (E) = \infty  .
  \eeq
\end{theorem}

The same is true for perturbations by boundary conditions: as a corollary of
Theorem \ref{kirsch1}, Kirsch \cite{Kir87} considered the Dirichlet Laplacian
on $\Lambda_L$ with an additional Dirichlet wall along the boundary $\partial
\Lambda_\ell$ of an arbitrary fixed cube $\Lambda_\ell \subset \Lambda_L$.  He
compared the eigenvalue counting function $N_{D,\ell}^{(L)}$ for this
operator, which is a direct sum of two Dirichlet Laplacians, to the one
obtained by placing Neumann boundary conditions along $\partial
\Lambda_\ell$. He concluded that $N_{N,\ell}^{(L)}(E) - N_{D,\ell}^{(L)}(E)$
has an infinite supremum over $L$, in the same way as in Theorem
\ref{kirsch1}. We remark that, similarly, this effect also shows up when
comparing $N_{D,\ell}^{(L)}$ to the eigenvalue counting function of the
Dirichlet Laplacian on $\Lambda_{L}$.

In the general case where $V_{0}$ is not identically zero, much less is
known. In fact, Kirsch's proof \cite{Kir87} of Theorem \ref{kirsch1} uses the
high degeneracy of the eigenvalues of $-\Delta_{L}$ to deduce the claimed
divergence of the SSF. In general, the perturbation $V_{0}$ removes this
degeneracy.

In \cite{Kir89}, Kirsch aimed at a complementary statement to
Theorem~\ref{kirsch1}, asking for finiteness of the SSF for energies outside
the bad countable set in \eqref{unbdd2}. The result he got, however, requires
$V$ to become smaller and smaller in norm when $L$ tends to infinity.

\begin{theorem}[\cite{Kir89}] \label{kirsch2} Let $d\in\NN$ and assume in
  addition to \eqref{ass} that $V_0 , V \in \mathrm{L}^{\infty}(\Rd)$.  Define
  $V_L := L^{-k} V$ for some arbitrary $k > d + 1$ and $\overline{H}_{1}^{(L)}
  := H_{0}^{(L)} + V_{L}$. Then, there is a set $S \subset \R $ of full
  Lebesgue measure such that for every $E \in S$, we have
  \beq\label{bdd1}
    \lim_{L \rightarrow \infty} \overline{\xi}_L (E) = 0.
  \eeq
  for the SSF $\overline{\xi}_{L}$ of the pair $(\overline{H}_{1}^{(L)},
  H_{0}^{(L)})$.
\end{theorem}

In our proof of a lower bound on the density of states for random
Schr\"odinger operators \cite{HiMu08}, we were also led to consider the
question of the boundedness of the finite-volume SSF. In the following theorem
we prove an almost-everywhere upper bound on the Ces\`aro mean of subsequences
of $\xi_{L}$. Our result naturally complements Kirsch's Theorem \ref{kirsch1},
which is commonly cited as an example of a pathological behavior of the SSF.

\begin{theorem}
  \label{kom-cor}
  Let $d\in\NN$ and assume \eqref{ass}. Then, for every sequence of lengths
  $(L_{j})_{j\in\NN} \subset \, ]0,\infty[$ with $\lim_{j\to\infty} L_{j} =
  \infty$ there exists a subsequence $(j_{i})_{i\in\NN} \subset\NN$ with
  $\lim_{i\to\infty} j_{i} =\infty$ such that for every subsequence
  $(i_{k})_{k\in\NN} \subset\NN$ with $\lim_{k\to\infty} i_{k} =\infty$ we
  have
  \begin{equation}
    \lim_{K\to\infty} \frac{1}{K} \sum_{k=1}^{K} \xi_{\wtilde{L}_{k}}(E) \le \xi(E)
  \end{equation}
  for Lebesgue-almost all $E\in\R$. Here we have set $\wtilde{L}_{k}
  :=L_{j_{i_{k}}}$ for all $k\in\NN$.
\end{theorem}

The assumption that $V$ has compact support could be relaxed. There have been
many works on the boundedness of finite-volume spectral shift functions for
Schr\"o\-dinger operators. Hundertmark, Killip, Nakamura, Stollmann and
Veseli\'c \cite{HuKiNa06} have obtained a bound on $ \int_\R \d E\; \xi_{L}(E)
\, f(E)$ for bounded, compactly supported functions $f$.  Kostrykin and
Schrader \cite[Ex.\ 4.2]{KoSc00} mention that their methods imply pointwise
boundedness of the sequence of Laplace transforms
$\big(\wtilde{\xi}_{L}(t)\big)_{L>0}$ for every fixed $t>0$. Combes, Hislop
and Nakamura \cite{CoHiNa01} proved an $\mathrm{L}^{p}$-bound on the SSF for
pairs of operators $(A,B)$ for which $C = B-A$ is in the Schatten-von Neumann
trace ideal $\mathcal{I}_{1/p}$, with $1 \leq p < \infty$. This was improved by
Hundertmark and Simon \cite{HuSi02} who obtained an optimal bound on the
$L^p$-norm of the SSF. Sobolev \cite[Sect.~4]{Sob93} showed continuity of the
infinite-volume SSF $\xi$ for pairs of Schr\"odinger operators with $V_{0}=0$
and proved a pointwise bound on $\xi(E)$ for sufficiently large energies
(there are more general results in \cite[Sect.~4]{Sob93} in an abstract
setting.) For the case of random Schr\"odinger operators on $L^2 (\R^d)$, it
is known that the expectation of the finite-volume SSF is pointwise bounded
\cite{CoHiKl07a, CoHiKl07b}.

The proof of Theorem~\ref{kom-cor} is deferred to Section~\ref{sec:komlos}. It
relies on a deep result of Koml\'os \cite[Thm.~1a]{Kom67} for
$\mathrm{L}^{1}$-bounded sequences. We infer this condition from the
non-negativity of $\xi_{L}$ and from vague convergence of the finite-volume
SSF towards the infinite-volume SSF. The latter is the content of

\begin{theorem}\label{main1}
  Let $d\in\NN$ and assume \eqref{ass}. Then, we have
  \beq\label{ssf-bound2}
    \lim_{L \rightarrow \infty} \int_\R \!\d E\; \xi_{L}(E) \, f(E) =
    \int_\R\!\d E \; \xi(E)\; f(E)
  \eeq
  for every continuous and compactly supported function $f \in C_c (\R)$ and
  for every indicator function $f=\Chi_{I}$ of some interval $I\subset
  \R$. In particular, for Lebesgue-almost all $E \in \R$ we have
  \beq\label{eq-ssf4}
  \lim_{\delta \downarrow 0} \lim_{L \rightarrow \infty}
    \frac{1}{\delta} \int_{E}^{E+\delta} \!\d E'\, \xi_{L}(E') = \xi(E).
  \eeq
\end{theorem}

\begin{remarks}
\item Geisler, Kostrykin, and Schrader \cite[Thm.~3.3]{GeKo95} proved that the
  distribution functions $\zeta_L (E) := \int_{- \infty}^E \!\d E\, \xi_L (E)$
  of the finite-volume spectral shift measures $\xi_L(E)\, \d E$ converge to
  the distribution function $\zeta (E)$ of the infinite-volume spectral shift
  measure in the case $d = 3$ and a real-valued measurable potential $V \in
  \ell^1 (\mathrm{L}^2)$, the Birman--Solomyak space.  In light of
  \cite[Prop~4.3]{HuLeMu01}, this proves vague convergence of the
  finite-volume spectral shift measures. The proof in \cite{GeKo95} uses
  Weyl-type high-energy asymptotics of the SSF (see \cite[Lemma~2.4]{GeKo95})
  that are not necessary for our proof.
\item Kirsch's result \eqref{unbdd1} shows that one cannot get rid of the
  energy smoothing in \eqref{eq-ssf4}, that is, the limits $\delta \downarrow
  0$ and $L\to\infty$ must not be interchanged. The best one could hope for is
  convergence Lebesgue-almost everywhere  of $(\xi_{L_{j}})_{j\in\NN}$ for
  sequences of diverging lengths. Theorem~\ref{kom-cor}  is a
  partial result in this direction based.
\item For the sake of concreteness, we mention an example of a toy family
  $(\zeta_{L})_{L >0}$ of functions
  \beq\label{ex2}
    \R \ni E \mapsto \zeta_L(E) := \left\{ \begin{array}{cc}
                   L, & E \in \{ \Q + [L]\} \\
                   0, & \rm{otherwise} \\
                 \end{array} \right.
  \eeq
  which captures the properties that are known for the family of spectral
  shift functions $(\xi_{L})_{L>0}$. In \eqref{ex2} we have written $[L] :=L
  \mod 1$ for the fractional part of $L$ in $[0,1[$. Indeed,
  $\lim_{L\to\infty} \zeta_{L}(E)$ does not exist for any $E\in\R$ and the
  corresponding suprema over $L$ are infinite as in \eqref{unbdd1} and
  \eqref{unbdd2}. The limits \eqref{ssf-bound2}, \eqref{eq-ssf4} are zero for
  the toy family. Note that, in addition, the sequence of functions
  $(\zeta_{L_{j}})_{j\in\NN}$ converges to zero Lebesgue-almost everywhere for
  every diverging sequence $(L_{j})_{j\in\NN}$ of lengths.
\end{remarks}

The Birman--Solomyak formula \cite{BiSo75, Sim98} is an important identity
relating the SSF to the perturbation potential $V$. We state it in the next
theorem and give a short proof of it in Section~\ref{sec:BS1}, valid under our
assumptions \eqref{ass}. Even though our proof of Theorem~\ref{main1} does not
rely upon the Birman--Solomyak formula, they are both 
related in spirit and are based on the Feynman--Kac formula. Let $\Chi_{B}$
denote the indicator function of the set $B \subset \R$.

\begin{theorem}\label{th:bs1}
  Let $H_0$ and $H_{1}$ be as above with potentials $V_0$ and $V$ satisfying
  \eqref{ass}, and let $H_{\lambda} := H_0 + \lambda V$ on
  $\mathrm{L}^{2}(\Rd)$ for $\lambda \in [0, 1]$. Then the (infinite-volume)
  SSF $\xi$ for the pair $(H_1, H_0)$ satisfies the Birman--Solomyak formula
  \beq\label{BS-rel}
    \int_B\!\d E \; \xi(E)=
    \int_0^1\!\d\lambda\; \tr[ V^{1/2} \Chi_B (H_\lambda) V^{1/2}]
  \eeq
  for every Borel set $B \subset \R$.
\end{theorem}

\begin{remarks}
\item
  We allow both sides of \eqref{BS-rel} to be $+\infty$. If $\sup B <
  \infty$, then
  \begin{equation}
    \tr[ V^{1/2} \Chi_B (H_\lambda) V^{1/2} ] = \int_{\R^{d}}\!\d x \, V(x) \,
    \Chi_{B}(H_{\lambda})(x,x) < \infty
  \end{equation}
  by \cite[Cor.~4.4]{Bri88}, the continuity of the integral kernel of the
  spectral projection (see \cite[Prop.\ 4.3]{AiSi82} or
  \cite[Thm.~B.7.1(d)]{Sim82}) and since $ V \in \mathcal{K}(\R^{d}) \subseteq
  \mathrm{L}^{1}_{\mathrm{unif,~loc}}(\R^{d})$ has compact support. Moreover,
  this trace is uniformly bounded in $\lambda \in [0,1]$.
\item
  The Birman--Solomyak formula provides another representation of the
  limiting measure $\xi (E)\,\d E$ in Theorem \ref{main1}.
  Using \eqref{BS-rel}, Eq.~\eqref{ssf-bound2} reads
  \begin{equation}
    \lim_{L \rightarrow \infty} \int_\R \!\d E\; \xi_{L}(E) \, f(E) =
    \int_0^1\!\d\lambda\tr[ V^{1/2} f (H_\lambda) V^{1/2}].
  \end{equation}
\item 
  Simon remarks that his more general Birman--Solomyak formula
  \cite[Thm.~4]{Sim98} includes the case of Schr\"odinger operators with
  slightly different conditions on the potentials than ours. For example, one
  may take $V_0$ and $V$ to be uniformly Kato class and $V \geq 0$ in the
  class $\ell_1 (L^2)$.
\end{remarks}


\section{Proof of Theorem~\ref{main1}}\label{sec-proof-1}

In this section we prove vague convergence of the finite-volume SSF in the
macroscopic limit. Our approach is very close to \cite{GeKo95} but does not
require knowledge of a Weyl asymptotics for high energies. Let $L,t > 0$. The
standard Feynman--Kac representation \cite{Sim79} of the heat kernel gives
\begin{align}\label{eq-ssf22}
  \wtilde{\xi}_L (t) :=& \int_\R \!\d E\; \e^{-tE} \xi_{L}(E)  \nonumber \\
  =& \frac{1}{t} \frac{1}{(2 \pi t)^{d/2}} \int_{\Lambda_L} \!\d x \;
  ~\E_{x,x}^{0,t} \left[ \Chi_{\Lambda_L}^t (b) \, \e^{- \int_0^t \d s \,
      V_0(b(s) )} \left( 1 - \e^{- \int_0^t \d s \, V(b(s)) } \right) \right]  .
\end{align}
Here, $\E_{x,y}^{0,t}$ denotes the normalized expectation over all Brownian
bridge paths $b$ starting at $x\in\R^{d}$ at time $s=0$ and ending at
$y\in\R^{d}$ at time $s=t$. The Dirichlet boundary condition is taken into
account by the cut-off functional $\Chi_{\Lambda_L}^t(b)$, which is equal to
one if $b(s) \in \Lambda_L$ for all $s \in [0, t]$, and zero otherwise.

First, we rewrite the term in parentheses in \eqref{eq-ssf22} as an integral
\begin{equation}
  \label{eq-integrate1}
  1 - \e^{- \int_0^t \d s \, V(b(s)) } =
  \int_0^1 \!\d\lambda  \left( \int_0^t \!\d s' \, V\big(b(s')\big) \right)
  e^{- \lambda \int_0^t \!\d s \, V(b(s))}
\end{equation}
over a parameter $\lambda \in [0,1]$. Then, we translate the Brownian paths
$b(s)$ to $b(s) + x$ and use Fubini's Theorem so that
\begin{equation}
  \label{eq-laplace2}
  \wtilde{\xi}_L(t) =  \frac{1}{t} \, \frac{1}{(2 \pi t)^{d/2}}
  \int_0^1 \!\d\lambda \, \E_{0,0}^{0,t} \big[ F_L(\lambda, t; b) \big],
\end{equation}
where
\begin{equation}
  \label{eq-laplace3}
  F_L(\lambda, t;b ) := \int_{\Lambda_L} \!\d x\,
  \Chi_{\Lambda_L}^t (b+x ) \left( \int_0^t
    \!\d s' \,  V\big(b(s')+x\big) \right) \e^{- \int_0^t \!\d s\, U_{\lambda}(b(s)+x) }
\end{equation}
and $U_{\lambda} := V_{0} + \lambda V$. Clearly, $F_{L}(\lambda, t;b) \ge 0$
is monotone increasing in $L$ for every Brownian bridge path $b$, every
$\lambda \in [0,1]$ and every $t>0$. Therefore, the Monotone Convergence
Theorem gives the pointwise limit
\begin{align}
  \label{mac-limit}
  \lim_{L\to\infty} \wtilde{\xi}_{L}(t)  =
  \frac{1}{(2 \pi t)^{d/2}}
  \int_0^1 \!\d\lambda  \int_{\R^{d}}\!\d x \int_{0}^{t}\frac{\d s}{t}
  \; \E_{x,x}^{0,t} \big[ V\big(b(s)\big) \, \cU_{t}(b)  \big]
\end{align}
for all $t>0$, where we have introduced the Brownian functional
\begin{equation}
  \label{U-def}
  \cU_{t}(b) :=\exp \bigg\{- \int_{0}^{t}\d s\, U_{\lambda}\big(b(s)\big) \bigg\}
\end{equation}
and used Fubini's Theorem. (We will see shortly that the limit
\eqref{mac-limit} is finite, which, a posteriori, will justify the
final interchange of integrations.)

Now we show that the limit \eqref{mac-limit} is equal to the two-sided
Laplace transform of the infinite-volume SSF. It is well-defined for
$t>0$, see Remark~\ref{rem-ssf}, and given by
\begin{align}
  \label{start-lap}
  \wtilde{\xi}(t) :=& \int_{\R}\d E\;  \e^{ -tE} \,
  \xi(E;H_{1},H_{0}) = - \int_{\R} \d E\;  \e^{ -tE} \,
  \xi\big( \e^{-tE}; \e^{-tH_{1}}, \e^{-tH_{0}}\big) \nonumber\\
 =& - \frac{1}{t} \int_{0}^{\infty}\!\d\eta \; \xi\big(\eta;
 \e^{-tH_{1}}, \e^{-tH_{0}}\big)
 = \frac{1}{t} \int_{\R}\!\d\eta \; \xi\big(\eta;
 \e^{-tH_{0}}, \e^{-tH_{1}}\big)
 \nonumber\\
 =& \frac{1}{t} \; \tr \big( \e^{-tH_{0}} - \e^{-tH_{1}} \big).
\end{align}
Here we have used the definition of the SSF in
Remark~\ref{rem-ssf}, and the last equality follows from Kre\u{\i}n's
trace formula \eqref{ssf-def-0}. The semigroup difference in the last
line of \eqref{start-lap} is trace class, cf.\ Remark~\ref{rem-ssf},
and possesses a continuous integral kernel. Thus, \cite[Thm.\
3.1]{Bri88} justifies the evaluation of the trace by an integral over
the diagonal of the kernel so that
\begin{align}
  \label{xi-lap}
  \wtilde{\xi}(t) &= \frac{1}{t} \, \frac{1}{(2 \pi t)^{d/2}}
  \int_{\R^{d}}\!\d x\; \E_{x,x}^{0,t} \left[
    \e^{- \int_0^t \d s \, V_0(b(s) )} \left( 1 - \e^{- \int_0^t \d s
        \, V(b(s)) } \right) \right]  \nonumber\\
  &= \frac{1}{(2 \pi t)^{d/2}}
  \int_0^1 \!\d\lambda  \int_{\R^{d}}\!\d x \int_{0}^{t}\frac{\d s}{t}
  \; \E_{x,x}^{0,t} \big[ V\big(b(s)\big) \, \cU_{t}(b)  \big],
\end{align}
where we have used \eqref{eq-integrate1} and Fubini's Theorem.
 We infer that
\begin{equation}
  \wtilde{\xi}(t) = \lim_{L\to\infty} \xi_{L}(t)
\end{equation}
for every $t>0$. In particular, the limit \eqref{mac-limit} is seen to be
finite (as was used earlier). So the claim \eqref{ssf-bound2} follows from
\cite{Fel71} for $f\in C_{c}(\R)$. But vague convergence of a sequence of
(unbounded) measures, which are tight at $-\infty$, implies pointwise
convergence of the corresponding distribution functions at points of
continuity of the limit, see e.g.\ \cite{Bau01} or
\cite[Prop~4.3]{HuLeMu01}. Thus, \eqref{ssf-bound2} also holds if $f$ is an
indicator function of some interval in $\R$. The statement \eqref{eq-ssf4} is
then a consequence of Lebesgue's Differentiation Theorem.  This completes the
proof of Theorem~\ref{main1}.


\section{Proof of Theorem~\ref{kom-cor}}\label{sec:komlos}

Since the SSF $\xi_{L}$ is non-negative for every $L>0$,
Theorem~\ref{main1} implies that for every sequence $(L_{n})_{n\in\N}$
of lengths, which is divergent to $+\infty$, we have
\begin{equation}
  \sup_{n\in\N} \int_{-j}^{j}\!\d E\; \xi_{L_{n}}(E) < \infty
\end{equation}
for every fixed $j\in\NN$, that is the sequence is norm bounded in
$\mathrm{L}^{1}([-j,j])$. Interpreting $(\xi_{L_{n}})_{n\in\NN}$ as a
sequence of uniformly distributed random variables on $[-j,j]$,
Koml\'os' Theorem \cite[Thm.~1a]{Kom67} ensures the existence of a
subsequence $(L_{n_{\nu}^{(j)}})_{\nu\in\N}$ of lengths and of a
function $\psi_{j} \in\mathrm{L}^{1}([-j,j])$ such that for every
further subsequence $\wtilde{L}_{k}^{(j)} := L_{n^{(j)}_{\nu_{k}}}$,
$k\in\NN$, the Ces\`aro limit
\begin{equation}
  \label{kom-j}
  \lim_{K\to\infty} \frac{1}{K} \sum_{k=1}^{K} \xi_{\wtilde{L}_{k}^{(j)}}(E)
  = \psi_{j}(E)
\end{equation}
exists for Lebesgue-a.e.\ $E\in [-j,j]$. Here we can assume without
restriction that $(n_{\nu}^{(j+1)})_{\nu\in\NN}$ is a subsequence of
$(n_{\nu}^{(j)})_{\nu\in\NN}$ for every $j\in \NN$. Below we show that
\begin{equation}
  \label{j-bound}
  \psi_{j} \le \xi
\end{equation}
holds Lebesgue-almost everywhere on $[-j,j]$ for
every $j\in\NN$. Therefore, given any subsequence $\wtilde{L}_{k} :=
L_{n_{\nu_{k}}^{(\nu_{k})}}$, $k\in\NN$, of the sequence
$(L_{n_{\nu}^{(\nu)}})_{\nu\in\NN}$, we get the asserted inequality
\begin{equation}
  \lim_{K\to\infty} \frac{1}{K} \sum_{k=1}^{K} \xi_{\wtilde{L}_{k}}(E)
  \le \xi(E)
\end{equation}
for Lebesgue-a.e.\ $E\in\R$.

It remains to establish \eqref{j-bound} for all $j\in\NN$. So fix
$j\in\NN$ and let $f \in C([-j,j])$ arbitrary, subject to $f \ge 0$
and $f(-j) = 0 = f(j)$. Then, the trivial extension $F$ of $f$ to $\R$
belongs to the positive cone of $C_{c}(\R)$.  We conclude from
\eqref{kom-j} and Fatou's Lemma that
\begin{align}
  \int_{-j}^{j} \!\d E\; f(E) \, \psi_{j}(E) &= \int_{\R}\!\d E\; F(E)
  \, \bigg( \lim_{K\to\infty} \frac{1}{K} \sum_{k=1}^{K}
  \xi_{\wtilde{L}_{k}}(E) \bigg)\nonumber\\
  &\le \int_{\R}\!\d E\; F(E) \, \xi(E) + \liminf_{K\to\infty}
  \frac{1}{K} \sum_{k=1}^{K} \mathcal{I}(k)\,,
\end{align}
where $\mathcal{I}(k) := \int_{\R}\!\d E\; F(E) \, [\xi_{\wtilde{L}_{k}}(E) -
\xi(E)]$.  Now, the vague convergence of Theorem~\ref{main1}
guarantees that for every $\varepsilon>0$ there exists
$k_{\varepsilon}\in\NN$ such that for all $k\ge k_{\varepsilon}$
we have $|\mathcal{I}(k)| \le \varepsilon$. This implies
\begin{equation}
  \bigg| \liminf_{K\to\infty} \frac{1}{K} \sum_{k=1}^{K} \mathcal{I}(k)
  \bigg|
  = \bigg| \liminf_{K\to\infty} \frac{1}{K} \sum_{k=k_{\varepsilon}}^{K} \mathcal{I}(k)
  \bigg| \le \varepsilon.
\end{equation}
Since $\varepsilon >0$ was arbitrary, we deduce
\begin{equation}
  \int_{-j}^{j} \!\d E\; f(E) \, [ \xi(E) - \psi_{j}(E) ] \ge 0
\end{equation}
for every non-negative $f\in C([-j,j])$ that vanishes at the boundary
points. But this yields \eqref{j-bound}.


\section{Proof of the Birman--Soloymak formula}\label{sec:BS1}

In this section we present a simple Feynman--Kac based proof of the
Birman--Solomyak formula \eqref{BS-rel}. Let us write $\mu(B)$ for the
right-hand side of \eqref{BS-rel}, which defines a Borel measure $\mu$ on
$\R$. We deduce from the spectral theorem and monotone convergence that for
every $t>0$ its two-sided Laplace transform $\wtilde{\mu}(t) :=
\int_{\R}\!\d\mu(E)\, \e^{-tE}$ is given by
\begin{align}
  \label{mu-lap}
  \wtilde{\mu}(t) &= \int_{0}^{1}\!\d\lambda \, \tr \big[ V^{1/2}
  \e^{-tH_{\lambda}} V^{1/2}\big] = \int_{0}^{1}\!\d\lambda \int_{\R^{d}}\!\d
  x \, V(x) \,
  \e^{-t H_{\lambda}}(x,x)      \nonumber\\
  &= \int_{0}^{1}\!\d\lambda \int_{\R^{d}}\!\d x \, V(x) \,
  \frac{\E_{x,x}^{0,t} \big[ \cU_{t}(b)\big]}{(2\pi t)^{d/2}}.
\end{align}
Finiteness of $\wtilde{\mu}(t)$ for $t>0$ and the second equality in
\eqref{mu-lap} hold because of \cite[Cor.~4.4]{Bri88}, the continuity of the
integral kernel of the semigroup \cite[Thm.~B.7.1(d)]{Sim82} and since $ V \in
\mathcal{K}(\R^{d}) \subseteq \mathrm{L}^{1}_{\mathrm{unif,~loc}}(\R^{d})$ has
compact support. The functional $\cU_{t}$ was defined in \eqref{U-def}.

Recall that the probability density $\rho_{x,x}^{0,t} (s;y)$, that the paths
of the Brownian bridge satisfy $b(s) = y \in \R^{d}$ for some $s \in ]0,t[$,
is given by
\begin{equation}
  \label{eq-density2}
  \rho_{x,x}^{0,t}(s;y) = (2 \pi t)^{d/2} \;
  \frac{\e^{-(x-y)^2/ (2s)}}{(2 \pi s)^{d/2}} \; \frac{ \e^{-(x-y)^2/[2(t-s)]}}{[2
    \pi (t-s)]^{d/2}}.
\end{equation}
The Markov property then amounts to the identity
\begin{align}
  \E_{x,x}^{0,t} \big[ \cU_{t}(b)\big] &= \int_{\R^{d}}\!\d y\,
  \rho_{x,x}^{0,t} (t-s;y) \, \E_{x,x}^{0,t} \big[\cU_{t}(b) \,\big| \, b(t-s)
  =y
  \big] \nonumber\\
  &= \int_{\R^{d}}\!\d y\, \rho_{x,x}^{0,t} (t-s;y) \, \E_{x,y}^{0,t-s}
  \big[\cU_{t-s}(b) \big] \, \E_{y,x}^{0,s} \big[\cU_{s}(b) \big]
\end{align}
for every $s\in ]0,t[$. Hence,
\begin{align}
  \label{mu-end}
  \int_{\R^{d}}\!&\d x\, V(x) \, \E_{x,x}^{0,t} \big[
  \cU_{t}(b)\big]
  \nonumber\\
  &= \int_{0}^{t} \frac{\d s}{t} \int_{\R^{d}}\!\d y
  \int_{\R^{d}}\!\d x\, \rho_{x,x}^{0,t} (t-s;y) \, \E_{y,x}^{0,s}
  \big[V\big(b(s)\big)\,\cU_{s}(b) \big] \,
  \E_{x,y}^{0,t-s} \big[\cU_{t-s}(b) \big] \nonumber\\
  &= \int_{0}^{t} \frac{\d s}{t} \int_{\R^{d}}\!\d y\,
  \E_{y,y}^{0,t} \, \big[V\big(b(s)\big)\,\cU_{t}(b) \big],
\end{align}
where the second equality relies on $ \rho_{x,x}^{0,t}(t-s;y) =
\rho_{y,y}^{0,t}(s;x)$ and, again, the Markov property. A comparison of
\eqref{mu-lap} and \eqref{mu-end} with \eqref{xi-lap} reveals that
$\wtilde{\mu}(t) = \wtilde{\xi}(t)$ for all $t>0$. Hence, \eqref{BS-rel}
follows from \cite{Fel71}.


\section{Appendix: Basics about the SSF}\label{sec-techni}

For the convenience of the reader, we collect some facts related to
the SSF in this appendix, see e.g.\ \cite{Yaf92}. First, we are
concerned with its definition in a more general setting. If $A_{0}$,
$A_{1}$ are self-adjoint operators on a Hilbert space $\cH$ and if
$A_{1} -A_{0}$ is trace class, then \cite[Thm.\ 8.3.3 and following
remarks]{Yaf92} $f(A_{1}) - f(A_{0})$ is trace class for every $f\in
C_{c}^{\infty}(\R)$ and the SSF $\xi \equiv \xi(\,\pmb{\cdot}\,; A_{1,},
A_{0})$ of the pair $(A_{1}, A_{0})$ is uniquely defined up to an
additive constant by Kre\u{\i}n's trace formula
\begin{equation}
  \label{ssf-def-0}
  \tr [f(A_{1}) -f(A_{0})] = \int_{\R}\!\d E \; f'(E) \,
  \xi(E).
\end{equation}
The constant can be chosen such that $\xi\in \mathrm{L}^{1}(\R)$. In
this case, we have the bound $\|\xi\|_{1} \le \|A_{1} -A_{0}\|_{\tr}$
in terms of the trace norm. We note that the behavior of $f$ outside
the union of the spectra of $A_{1}$ and $A_{0}$ is irrelevant for both
sides of \eqref{ssf-def-0} and that \eqref{ssf-def-0} does also hold
for $f$ being the identity.

This definition of the spectral shift function
$\xi(\pmb{\cdot};A_{1,}, A_{0})$ can be extended to a pair of
self-adjoint operators $(A_{1}, A_{0})$ for which it is only known
that $\varphi(A_{1}) -\varphi(A_{0})$ is trace class for some function
$\varphi \in C^{2}(\Omega)$, where $\Omega$ is a (possibly infinite)
interval containing the union of the spectra of $A_{0}$ and $A_{1}$
and $\varphi$ is bounded and strictly monotone on $\Omega$. In this case
we set
\begin{equation}
  \label{ssf-def-1}
  \xi(E; A_{1}, A_{0}) := \mathrm{sign}\big(\varphi'(E)\big)\; \xi\big
  (  \varphi(E) ;\varphi(A_{1}) , \varphi(A_{0})\big)
\end{equation}
for all $E\in \Omega$. The SSF on the right-hand side of
\eqref{ssf-def-1} is determined by \eqref{ssf-def-0} and it is
independent of the choice of $\varphi$ within the allowed class of
functions (invariance principle) \cite[Sect.\ 8.11]{Yaf92}. Moreover,
we have the estimate
\begin{equation}\label{eq:ssf-bd1}
  \int_{\Omega}\!\d E\; |\xi(E; A_{1}, A_{0})|\, |\varphi'(E)| \le \|
  \varphi(A_{1}) -\varphi(A_{0})\|_{\tr} .
\end{equation}

Now we return to the situation of Schr\"odinger operators as in the main text.

\begin{remark}
  \label{rem-ssf}
  Let $d\in\NN$ and assume \eqref{ass}. Then, $\e^{-tH_{0}} -\e^{-tH_{1}}$ is
  trace class for every $t>0$, see e.g.\ \cite[Remark after Thm.~1]{HuKiNa06},
  and the SSF for the pair $(H_{1}, H_{0})$ is defined by \eqref{ssf-def-1}
  with $\varphi(E) := \e^{-tE}$ for $E\in\R$. It follows that
  $\xi(\,\pmb{\cdot}\,, H_{1}, H_{0}) \in \mathrm{L}^{1}_{\mathrm{loc}}(\R)$, the
  integral corresponding to (\ref{eq:ssf-bd1}) is finite, and that its Laplace
  transform $\wtilde{\xi}(t)$, see \eqref{start-lap}, exists for every $t>0$.
\end{remark}

\begin{acknowledgement}
  The authors thank Fran\c{c}ois Germinet for repeated kind hospitality at the
  Universit\'e de Cergy-Pontoise, France.
\end{acknowledgement}


\end{document}